\documentclass[twocolumn]{wiley-article}

\usepackage{siunitx}

\usepackage{color}
\usepackage{amsmath}
\usepackage{amssymb}
\usepackage{graphicx}
\usepackage{placeins}
\usepackage{enumitem}
\usepackage[english]{babel}

\usepackage{tabto}
\usetikzlibrary{shapes,arrows}
\usetikzlibrary[pgfplots.groupplots]
\usetikzlibrary{calc}
\usepackage{pgfplots}
\usetikzlibrary[pgfplots.groupplots]
\usepgfplotslibrary{fillbetween}
\usepackage{filecontents}
\usetikzlibrary{external}

\newcommand{\figWidth}{0.6\columnwidth}

\usepackage[pdftex,bookmarksnumbered,bookmarks=true,plainpages=false]{hyperref}
\hypersetup{colorlinks=true}
\hypersetup{linkcolor=blue}
\hypersetup{citecolor=blue}
\hypersetup{urlcolor=blue}

\usepackage{textcomp}
\usepackage{tabularx,colortbl}
\usepackage{todonotes}
\usepackage[]{algorithm2e}
\usepackage{listings}

\usepackage{setspace}
\newcommand{\zweizeilig}{}

\papertype{Original Article}
\paperfield{Journal Section}

\title{Simulation of reactive flows\\ using particle methods}

\author{Sebastian M\"uhlbauer\!}
\author{\!\!Severin Strobl}
\author{\!\!Thorsten P\"oschel}

\affil{Institute for Multiscale Simulation, Friedrich-Alexander-Universit\"at Erlangen-N\"urnberg, Cauerstra{\ss}e 3, 91058 Erlangen, Germany} 

\corraddress{Thorsten P\"oschel, Institute for Multiscale Simulation, Friedrich-Alexander-Universit\"at Erlangen-N\"urnberg, Cauerstra{\ss}e 3, 91058 Erlangen, Germany}
\corremail{thorsten.poeschel@fau.de }

\fundinginfo{DFG, SFB, ZISC, FPN}

\runningauthor{Sebastian M\"uhlbauer et al.}
\date{\today}
\paperreceived{\today}
\begin{document}
 \zweizeilig

\begin{frontmatter}
\maketitle

\begin{abstract}
We describe a new computational method for the numerically stable particle-based   simulation of open-boundary flows, including volume conserving chemical reactions. The novel method is validated for the case of heterogeneous catalysis against a reliable reference simulation and is shown to deliver identical results while the computational efficiency is significantly increased.
\end{abstract}
\end{frontmatter}

\section{Introduction}

\label{intro}

In particle-based fluid simulation methods, pressure boundary conditions, also called {\em open boundaries} are notoriously problematic since either the inflow into the simulation domain or the outflow out of the domain is not known {\`a priori}. Instead, it depends on the difference of the external pressure given as a boundary condition and the internal pressure which is part of the solution of the hydrodynamic problem itself. Thus, we have the peculiar situation that the quantity we can control at the boundary in particle-based fluid simulation, namely the material flux, is unknown.

A prototypical problem is a water pipe which is fed from a constant pressure source at one side and emptied against atmospheric air pressure at the other side. According to the paradigm of particle-based hydrodynamics, quasi-particles enter and leave the pipe at both sides in quantities given by the rules of hydrodynamics. The problem of the computational method of particle-based hydrodynamics is to determine the frequency at which particles are inserted into or extracted from the simulation domain as well as the velocities of the inserted particles. If the hydrodynamic fields of pressure, temperature and flow velocity would be know at the boundaries, that is, if  Dirichlet boundary conditions for all relevant macroscopic fields are assumed, the injected and extracted particles can be drawn from a probability distribution function \cite{Garcia2006}.

From a theoretical point of view, a pressure boundary is established by a reservoir, therefore, another well-established method to model open boundaries in particle-based fluid simulations is to attach a fluid reservoir to each boundary surface containing particles at local equilibrium \cite{Garcia1999,Macrossan2003,Lilley2003,Tysanner2005}. However, if we take the concept of a reservoir literally, it has to be of infinite size or, at least, of a size much larger than the size of the simulation domain, leading to very large computational costs of this approach. In practically all cases where Dirichlet boundary conditions for at least one of the hydrodynamics fields are not appropriate, the simulation of the boundaries is problematic \cite{Gatsonis2013}.

In the most basic case, one can assume that no particles enter the simulation domain at the outlet boundary, which corresponds to an outflow without backflow. This assumption is valid for flows into vacuum and can also be justified in cases where the flow velocity at the outlet is large compared to the thermal velocity of the individual particles, that is, for supersonic flows. Examples are  micro-scale propulsion systems for spacecraft \cite{Sun2009,Darbandi2013}. For the majority of engineering problems, however, a particle backflow is required at the outlet. Considering, for example, micronozzles in microelectromechanical systems, typically a finite pressure is assumed at the outlet \cite{Alexeenko2006,Morinigo2010,Sebastiao2014}.

If the above approach is not justified and not all fields are known {\em \`a priori}, more complex boundary conditions have to be applied \cite{Gatsonis2013,Ikegawa1990,Nance1998,Whitfield1984,Cai2000,Wang2004,Farbar2014,Chamberlin2007,Thompson1990}, however, numerical stability is not guaranteed for such models. For example, boundary conditions for which the velocity field at the boundaries is not known, while pressure and temperature are given, tend to induce instabilities, especially for small Mach number \cite{Cai2000,Wang2004}.

If the considered domain is periodic in flow direction, and the Mach number is small, i.e., the fluid is incompressible, the difficulties elaborated above can be circumvented by applying periodic boundaries \cite{Metropolis1953,Alder1955}. Further, an external acceleration can then be used to model a pressure gradient and, thus, drive the flow \cite{Alaoui1992}.  This procedure, however, enforces the periodicity on all involved fields. While this is frequently justified for density, velocity, and temperature, it is not applicable for the concentration field in reactive flow, for example, for modeling catalytic converters.

In the current paper we describe a novel computational method for the particle-based simulation of systems where the fields of  density, velocity, temperature, and pressure are periodic while  the concentration field may develop a discontinuity at the interface. The modus operandi of this partially periodic boundary does not depend on the particular simulation method used. It works equally for all mesoscopic, particle-based simulation methods. The simulation approach is validated for the case of heterogeneous catalysis and the enhanced computational efficiency is demonstrated.

\section{Method}

The proposed boundary condition acts on pairs of the form {\em reactant species}/{\em product species}. For this concept to work, the chemical reaction has to be volume preserving, i.e., the number of product particles must equal the number of reactant particles. 
For illustration, we consider the most simple volume preserving chemical reaction: $ A \rightarrow B $. For more complex reactions, $ A_1 + A_2 + ... + A_n \rightarrow B_1 + B_2 + ... + B_n $, the following concept acts upon all pairs $ \left(A_1 ,B_1\right) $, \dots, $\left( A_n, B_n \right)$. 
Note that in general, the mass of a particle changes when undergoing a reaction $A\to B$, that is, $m_A \neq m_B $. In this case, to keep the temperature continuous, the thermal velocity of the particle has to be adjusted whenever its state changes, due to a reaction or passing the periodic boundary.

\begin{figure}[htb]
	\centering
	\includegraphics[width=\figWidth]{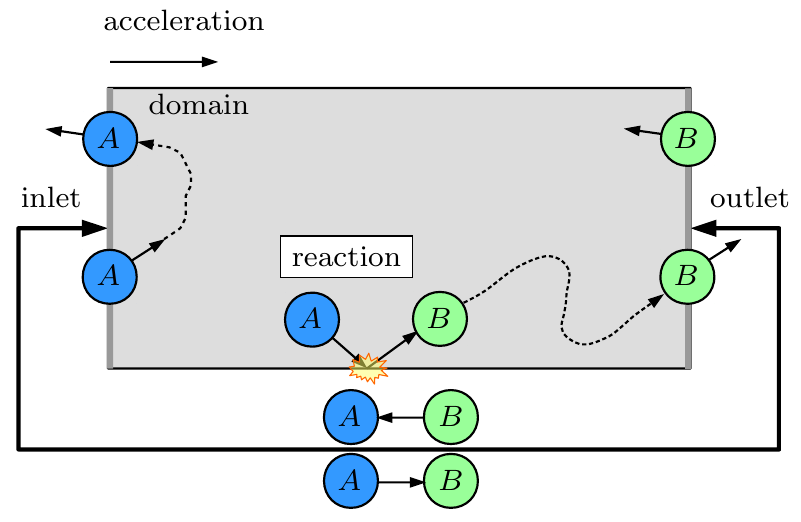}
	\caption{\zweizeilig Illustration of the working principle of the boundary model: Particles of type $A$ enter the system at the left side (inlet). Reactions $ A \rightarrow B$ may take place in the interior of the system. In the illustration, we assume that reactions take place at the lower wall which may be coated by a catalyst. Particles of type $A$ and $B$ leave the system at the right boundary (outlet). For further explanation see the text.}
	\label{fig:SketchReset}
\end{figure}

When a particle of type $A$ undergoes a chemical reaction, it changes its type to $B$. Fig. \ref{fig:SketchReset}
illustrates the working principle of the boundary model. Particles of type $A$ enter the system at the left side (inlet). Reactions $ A \rightarrow B$ may take place in the in interior of the system. Particles of type $A$ and $B$ leave the system at the right boundary (outlet). We call the direction inlet $\to$ outlet the mean flow direction, driven by a pressure difference between the inlet and the outlet sides. If the distance between inlet and outlet is large and/or the reaction rate is large, at the outlet there appear almost exclusively particles of type $B$. Na\"{\i}vely thinking, these particles can be reset to type $A$ and re-inserted into the inlet to close the periodical boundary conditions. This is, however, incorrect: Albeit the particles move preferably in the main flow direction, individual particles can also cross the boundary at the inlet side and would have to be inserted at the outlet side. This would lead to a substantial concentration of particles of type $A$ at the outlet side which contradicts our model assumption above. In the described system this error could be repaired by performing a symmetric transformation $A\to B$ for particles crossing the periodic boundary against the main flow direction. This is, however, not applicable if in the physical system a non-negligible fraction of particles are of type $A$ close to the outlet boundary. Thus, for the important case of incomplete reaction, we need a more sophisticated simulation method. As a further complication we have to take into account that in dependence on the system size, the reaction rate, and the flow velocity, a particle may:
\begin{enumerate}
  \renewcommand{\labelenumi}{({\it\roman{enumi}}) }
\item cross the period in the main flow direction several times before, on average, it undergoes a reaction $A\to B$, or,
\item cross the period against the main flow direction one or more times due to its individual velocity. The particle can be either of type $A$ or $B$.
\end{enumerate}
For a reliable simulation algorithm, these cases have to be taken into account. Essential for the description is the {\em reset} ($A\to B$ or $B\to A$) of the particle in the process of crossing the periodic boundary.

\section{Algorithm}
\label{sec:algorithm}

For the description of the algorithm, \texttt{species} denotes the current state of a particle, that is, \texttt{species}$\,\in \{A,B\}$. \texttt{product\_species} denotes the species of a particle after a reaction. In our example \texttt{product\_species}$\,= B$. The variable \texttt{target\_species} holds the state the particle assumes when crossing the periodic boundary. The value of  \texttt{target\_species} depends on the history of the particle,  characterized by the variable \texttt{passes} which counts how often a particle has passed the simulation domain since the last reaction. Passes in main flow direction count positively, while passes against flow direction count negatively. For simplicity, the initial state for all particles is assumed to be:\\
\texttt{species} := A,\\
\texttt{product\_species} := B,\\
\texttt{target\_species} := A,\\
\texttt{passes} := 1.

The following three events have to be considered:
\begin{enumerate}
  \renewcommand{\labelenumi}{({\it\roman{enumi}}) }
  \renewcommand{\theenumi}{{\it\roman{enumi}}}
\item \label{it:reaction}
  A particle undergoes a {\em reaction}:\\
  In case a particle undergoes a chemical reaction, \texttt{target\_species} assumes the value \texttt{species} if this is the first reaction since the last reset (\texttt{passes} $ \neq $ 0). This ensures that for serial reactions $ A \rightarrow B $ and $ B \rightarrow C $ the \texttt{target\_species} is not changed to $ B $, but remains $ A $, even after the second reaction, $ B \rightarrow C $, has taken place. The type of a particle, \texttt{species}, assumes the value \texttt{product\_species}, due to a reaction and \texttt{passes}\ =\ 0, indicating that the particle did not cross the periodic boundary after performing a reaction.

  \begin{algorithm}
  \If{\normalfont (\texttt{passes} $ \neq $ 0)}{
    \hspace*{2.5mm} \texttt{target\_species} := \texttt{species}\;
  }
  \texttt{species} := \texttt{product\_species}\;
  \vspace*{1mm}

  \texttt{passes} := 0\;
  \end{algorithm}

\item\label{it:reset}
  A particle crosses the periodic boundary in main flow direction (outlet $\to$ inlet):\\
  Here we have to distinguish two cases: If the particle did not cross the periodic boundary since the last reaction indicated by \texttt{passes}\ =\ 0, its state is reset to the initial value while  \texttt{target\_species} assumes the value of the current state. Additionally, its counter for the number of passes is incremented. This means, when a particle crosses the periodic boundary against the main flow, we have to restore its state as it was just before the previous crossing in direction of the main flow, see item \ref{it:rereset}. Moreover, if the number of passes is zero, that is, the particle is not reactive (it just had a reaction), after the transition it has to be set again to the reactive state. To achieve this behavior, \texttt{passes} is unconditionally incremented for each particle crossing the periodic boundary in the main flow direction.
  \begin{algorithm}
  \If{\normalfont (\texttt{passes} = 0)}{
    \hspace*{2.5mm} swap (\texttt{target\_species}, \texttt{species})\;
  }
  \texttt{passes} := \texttt{passes} + 1\;
  \end{algorithm}

\item \label{it:rereset}
  A particle crosses the periodic boundary opposite to the main flow direction (inlet $\to$ outlet). In this case, we again have to distinguish three cases:
  \begin{enumerate}
  \item \texttt{passes}\ =\ 1 is the state which is assumed when a particle that has undergone a reaction just crossed the periodic boundary in direction of the main flow. Thus, when transiting in opposite direction, it should assume the state it had after undergoing a reaction, that is, \texttt{passes}\ =\ 0 and the values of \texttt{target\_species} and \texttt{species} should be exchanged.

  \item \texttt{passes}$\ >\ $1 indicates that the particle has passed the period in the flow direction at least once without undergoing a reaction. In this case, when passing the boundary in direction opposite to the main flow, \texttt{species} remains unchanged and the counter \texttt{passes} is decremented.

  \item \texttt{passes}=0 indicates that the particle has passed the period at least once in direction opposite to the flow. In this case all variables remain unchanged.
  \end{enumerate}

   \begin{algorithm}
  \If{\normalfont (\texttt{passes} = 1)}{
    \hspace*{2.5mm} swap (\texttt{target\_species}, \texttt{species})\;
  }
  \If{\normalfont (\texttt{passes} $ > $ 0)}{
    \hspace*{2.5mm} \texttt{passes} := \texttt{passes} - 1\;
  }
  \end{algorithm}

Obviously, the behavior arising when a particle crosses the periodic boundary in the direction of the main flow and opposite to it is not symmetric since $ \texttt{passes} $ cannot assume negative values, independent of how many times a particle crossed the boundary in direction opposite to the main flow. More specifically:
\begin{itemize}
  \item A particle, having passed $ N $ times through the domain along the main flow direction, has to pass the boundary in the other direction $ N - 1 $ times before it assumes again an inert state.
  \item A particle, having passed $ N $ times through the domain against the main flow direction, has to pass the boundary in the other direction only once to assume a reactive state.
\end{itemize}
To understand this asymmetry, we consider a $t$-step random walk of a particle in one-dimensional space, which is performed multiple times with identical initial conditions. For increasing number of repetitions, the mean displacement, $\left< X_t \right>$, approaches zero while the mean squared displacement, $ \left< X_t^2 \right>$, does not. This means, that random walkers tend to move away from the initial position symmetrically in both directions. In relation to our system with periodic boundary conditions, the actual number of passes of each particle in principle follows a one-dimensional random walk, biased by the flow field. This is, however, not true for the variable \texttt{passes}, since each reaction causes \texttt{passes}\ =\ 0 which introduces an asymmetry. If \texttt{passes} had no lower bound, its mean value, $ \left<\texttt{passes} \right>$, would approach $ -\infty $. Thus, after some time, virtually all particles would cease reacting since only for $ \texttt{passes} \in \{0,1\} $ the state of the particles changes when passing the periodic boundary.
\end{enumerate}

\section{Validation}

\subsection{System setup}
\begin{figure}[htb]
	\centering
	\includegraphics[width=\figWidth]{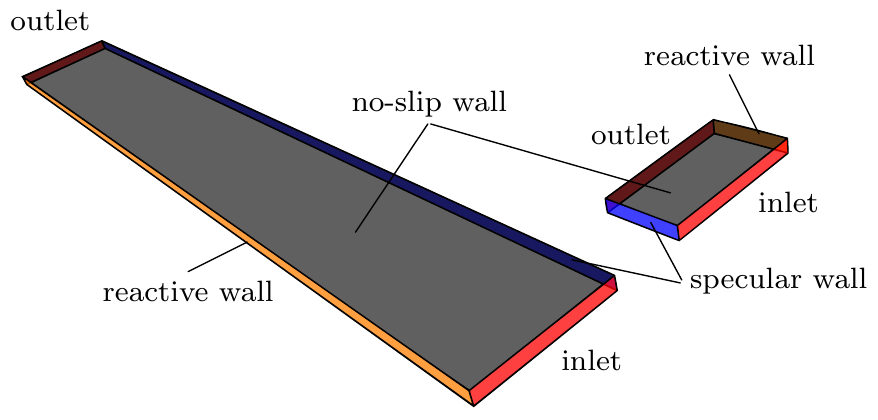}
	\caption{\zweizeilig For validation we compare the simulation results for a system of size $40\times 4 \times 320$ with the results for a system of size $40\times 4 \times 20$. In the first case near to the outlet virtually all particles have experienced a reaction, thus, the simulation is plain. In contrast, for the second system this is not the case such that the behavior at the periodic boundary is essential for the simulation.}
	\label{fig:TestSetup}
\end{figure}

To demonstrate the validity of the algorithm, we simulate heterogeneous catalysis, that is, the reactions are bound to a {\em reactive wall} as sketched in Fig. \ref{fig:TestSetup}.
In the mass transfer limited regime, a characteristic concentration profile is expected which can be used to verify the algorithm. For verification we will compare the mole fraction profiles in two different channels which are identical except for their size along the main flow direction. These two channels together with the applied boundary conditions are depicted in Fig. \ref{fig:TestSetup}. The longer channel, denoted as {\em reference}, consists of $ 40 \times 4 \times 320 $ cubic cells, while the shorter one
comprises only $ 40 \times 4 \times 20 $ cubic cells. The cell width, $a = 1$, is the same for both channels. The mole fraction of species $i$ is defined as
\begin{equation}
  \label{eq:concentration}
  x_i = \frac{n_i}{n_{\text{tot}}} \, ,
\end{equation}
where ${n}_i$ is the number density of species $i$ and ${n}_\text{tot}$ is the total number density. For the case of the long channel the reaction is nearly complete close to the outlet, that is, the local concentration of species $A$ vanishes, $x_A^\text{outlet}\to 0$. In this limit, the periodic boundary can be closed in the simplified version as explained above. This is different for the short channel, where close to the outlet a significant concentration of species $A$ is present, that is, the reaction is incomplete. Therefore the more sophisticated algorithm for modeling of the boundary conditions explained in the previous section is needed. Assuming the suggested boundary conditions work as expected, the profile of mole fraction of the short channel should be identical to the first part of the reference channel, that is, the resemblance of both profiles may be considered as a measure for the correctness of the algorithm.

The choice of the specific method for the simulation of the gas dynamics is irrelevant for the validation of the boundary model; here we use Direct Simulation Monte Carlo (DSMC) \cite{Bird1994} with mean free path, $\lambda = 1$, equal to the cell width, $a$.

\subsection{First-order reaction}

We consider the surface reaction $ A \rightarrow B $, as sketched in Fig. \ref{fig:SketchReset}. When a particle of type $A$ touches the catalytic surface, it changes its type to $ B $. We initialize both domains with $10$ particles of type $A$ and $10$ particles of type $B$ per simulation cell. The reactant mole fraction is, thus, homogeneously $x_A=0.5$. In both cases, inlet and outlet are linked via the periodic boundary conditions as described above.

\begin{figure}[htb]
	\centering
	\includegraphics[width=\figWidth]{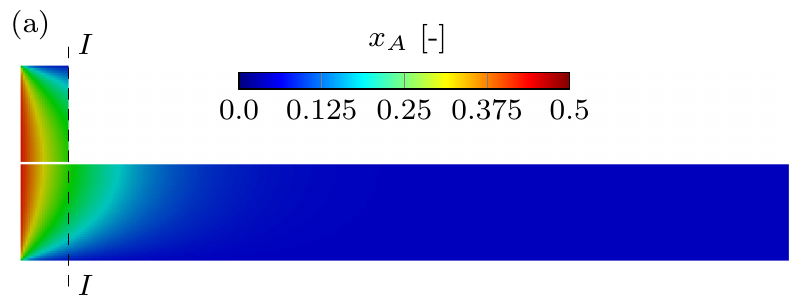}
	\includegraphics[width=\figWidth]{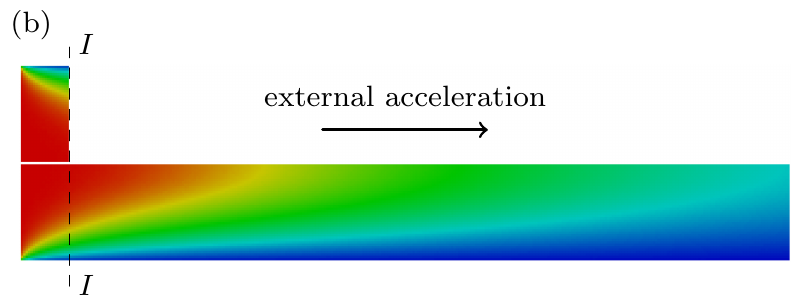}
	\caption{\zweizeilig Profiles of reactant mole fraction, $x_A$, along the channels for the case of (a) pure diffusion and (b) a flow driven by an external acceleration. In both cases the upper part of the image shows the short channel using the novel boundary condition and the lower part shows the reference channel.}
	\label{fig:Concentration}
\end{figure}

First we consider the case of pure diffusion, that is, there is no driving force which would impose a main flow direction. In the second instance, the flow is driven by an external acceleration, equivalent to a pressure gradient along the length of the channel. The superficial and maximum velocity in the channel are $0.14 \, c_s $ and $0.19 \, c_s $, respectively, where $c_s$ is the speed of sound. The resulting mole fraction fields are shown in Fig. \ref{fig:Concentration}.

\begin{figure}[htb]
	\centering
	\includegraphics[width=\figWidth]{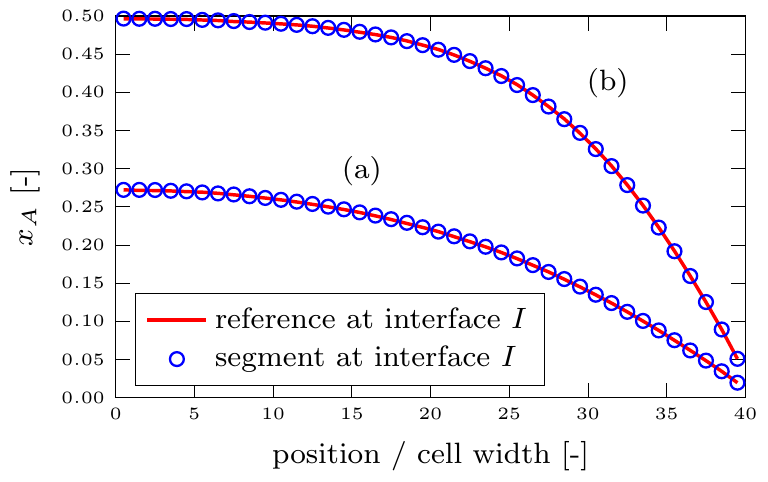}
	\caption{\zweizeilig Profiles of the reactant mole fraction profiles right at the outlet of the short channel, marked by  $I$ in Fig. \ref{fig:Concentration}. The labels, (a) and (b), refer to the two cases shown in Fig. \ref{fig:Concentration}.}
\label{fig:ConcentrationGraph}
\end{figure}

For both cases -- with and without external forcing -- the reactant mole fraction in the short channel does not reveal any noticeable difference compared to the reference channel. For a more quantitative comparison, Fig. \ref{fig:ConcentrationGraph} shows the profile of the reactant mole fraction in direction perpendicular to the flow direction, close to the outlet of the short channel, indicated by $I$ in Fig. \ref{fig:Concentration}.
Even right at the short channel's outlet, at position $I$, the obtained mole fraction profiles for the short channel coincide exactly with the curves obtained for the reference channel. $I$ is the most critical position in the simulation since it is heavily affected by the boundary condition model. From the good agreement, we conclude that our model of the  periodic boundary provides the correct mole fraction profile at the outlet.

\subsection{Second-order reaction}
\begin{figure}[htb]
	\centering
	\includegraphics[width=\figWidth]{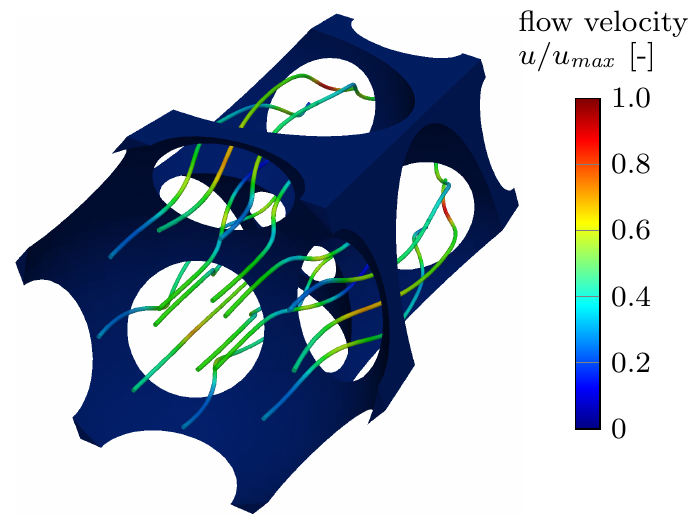}
	\caption{\zweizeilig Streamlines for gas flow in an open-cell porous foam structure modeled as inverse sphere packing using constructive solid geometry (CSG) \cite{Strobl2020}.}
\label{fig:FoamFlow}
\end{figure}

As a second example for validation and to emphasize the potential of our approach, we simulate a more challenging example, namely the low-temperature water-gas shift reaction \cite{Ayastuy2005}
$$
\text{CO} + \text{H}_{2}\text{O} \rightarrow \text{CO}_{2} + \text{H}_{2}
$$
in an open-cell porous foam structure, modeled by an inverse sphere packing \cite{Smorygo2011}. Figure \ref{fig:FoamFlow} shows the system. Here, the gas dynamics was simulated using  isotropic stochastic rotation dynamics (iSRD) \cite{Muehlbauer2017}.
Again, the flow is driven by an external acceleration, resulting in the superficial and maximum velocities $ 0.10 \, c_s $ and $ 0.30 \, c_s $, respectively. For the considered setup, the pore size Reynolds number is
\begin{equation}
  \label{eq:Rey}
  \text{Re} = \frac{\text{superficial velocity} \times \text{pore size}}{\text{kinematic viscosity}} = 26 \, ,
\end{equation}
while the Schmidt number is $0.77$. The parameters, in particular the local reaction rate, are chosen such that the system operates in  the mass transfer limited regime, that is, practically all encounters of $ CO $ and $ H_{2}O $ at the surface lead to a reaction. The inflow mole fractions of the reactants are $ x_\text{CO} = 3 \, \% $ and $ x_{\text{H}_{2}\text{O}} = 26 \, \% $.

In this case, where the reactants' inlet concentrations differ, the change of the particle type when crossing the periodic boundary opposite to the main flow direction (event \ref{it:rereset} in Sec. \ref{sec:algorithm}) must be synchronized, such that the stoichiometric ratio of the reactants is preserved, on average. For second-order reactions with differing reactant concentrations at the inlet, this can be achieved by the following procedure: In each cell located at the periodic boundary, we count the number of type changes per species due to particle crossings against the flow direction. While for the rare reactant species, $CO$, the type is adapted unconditionally, type changes for the abundant species, $H_{2}O$, are only performed if the rare species, has experienced a larger or equal number of corresponding changes.

\begin{figure}[htb]
	\centering
	\includegraphics[width=\figWidth]{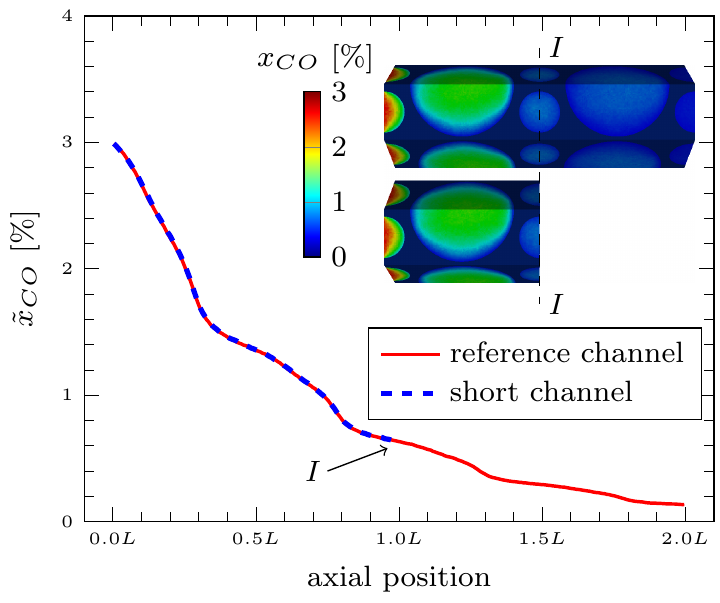}
	\caption{\zweizeilig Particle flow rate of $CO$ over total particle flow rate (see Eq. \eqref{eq:concentration2}) for an open-cell porous foam structure. The inset shows the mole fraction of $CO$ in the two considered systems}
\label{fig:FoamConc}
\end{figure}

\begin{figure}[htb]
	\centering
	\includegraphics[width=\figWidth]{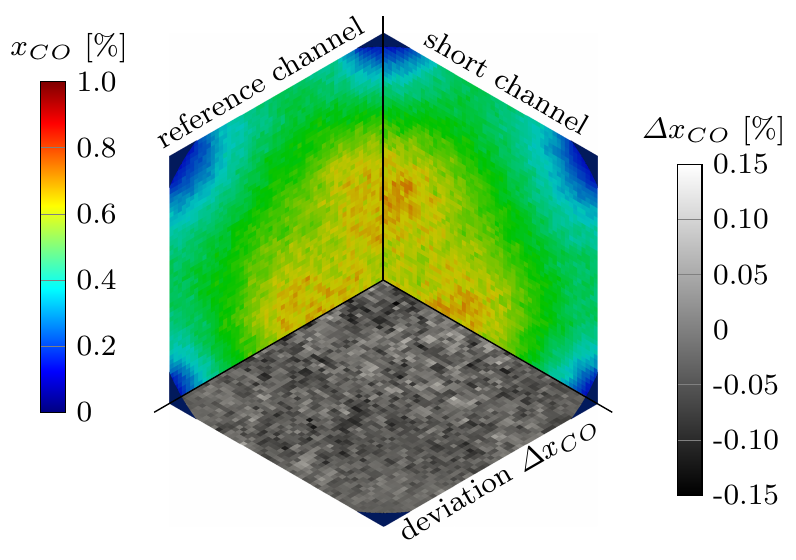}
	\caption{\zweizeilig Mole fraction profile, $x_\text{CO}$ at position $I$ for the hexagonal cross section of the channels, that is, perpendicular to the flow direction. According to the symmetry of the system (see Fig. \ref{fig:FoamFlow}), we subdivide the cross section into three identical parts. The top right: $x_\text{CO}$ as obtained from the simulation of the short channel. Top left: $x_\text{CO}$ profile for the reference channel. Bottom: difference between the two profiles.}
	\label{fig:FoamConc2}
\end{figure}

Figure \ref{fig:FoamConc} illustrates the consumption of $CO$ along the axial direction of the foam structure.
Instead of the mole fraction, we plot the ratio between the particle flow rates,
\begin{equation}
  \label{eq:concentration2}
  \tilde{x}_i = \dfrac{\int_\text{S} \, {n}_{i} \, u \, \text{d}s}{\int_\text{S} \, {n}_{\text{tot}} \, u \, \text{d}s} \, ,
\end{equation}
where the integral runs over slices with constant axial coordinate. The velocity component in axial direction is denoted as $u$, while $ds$ is the surface element. Again the two compared simulation setups shown as inset in Fig. \ref{fig:FoamConc} differ only in the length of the domains, with the reference channel being formed by two unit cells. Any deviations of the results for the short channel in comparison with the reference channel due to imperfect handling of the boundary conditions should be apparent in the concentration field close to the position $I$. Figure \ref{fig:FoamConc2} shows the mole fraction profile at position $I$ for the hexagonal cross section of the channels, that is, perpendicular to the flow direction.
According to the symmetry of the system (see Fig. \ref{fig:FoamFlow}), we subdivide the cross section into three identical parts showing the profiles of $x_\text{CO}$ as obtained from the simulation of the short channel (top right) and from the reference channel (top left). From the lower part, which shows the difference between both profiles,
$\Delta x_{CO} = x_{CO}^\mathrm{short} - x_{CO}^\mathrm{reference}$, we can see that the results differ by not more than about 0.15\% which can be attributed to thermal fluctuations. Thus, again we find excellent agreement between the short channel and the reference channel.

\section{Conclusion}

We presented an algorithm for the particle-based simulation of open boundaries in reactive flows through geometrically periodic domains. Our simulation domain is equipped with periodic boundary conditions for the fields of density, flow velocity, and temperature, but reveals a discontinuity with respect to the fields of concentration of the reactants. The concentration field emerging due to the proposed boundary condition corresponds to a channel, which is periodically continued at the outlet, and has a homogeneously mixed inflow. Our boundary condition is not as general as the characteristic boundary condition suggested in \cite{Gatsonis2013}, however, it is numerically stable. In contrast to \cite{Gatsonis2013}, no partial differential equations have to be solved for the inlet and outlet. Alternatively, a conceptually simple approach is to choose the system large enough so that the reaction can be safely assumed to be complete, i.e., the reactants' concentration is known at the outlet. Compared to this, the approach described in this work can reduce the numerical cost to a fraction. Additionally, for some applications, homogeneous velocity and concentration fields may be desired both at inlet and outlet, for example to mimic reservoirs allowing for unknown average concentrations at the outlet. This can be achieved by inserting the particles at random positions, when crossing the periodic boundary in both directions.

\section*{Acknowledgements}

We thank Prapanch Nair for sharing his expertise in fluid dynamics and the German Research Foundation (DFG) for funding through SFB\,814, ZISC, and, FPS.

\bibliographystyle{aichej}
\bibliography{reactiveFlow}
\end{document}